\documentclass[12pt]{article}

\def\bphi{\mbox{\boldmath $\phi$}}

\def\balpha{\mbox{\boldmath $\alpha$}}

\def\bpsi{\mbox{\boldmath $\psi$}}
\def\bgamma{\mbox{\boldmath $\gamma$}}
\def\bchi{\mbox{\boldmath $\chi$}}

\input{amssymb.sty}

\def\bbc{{\mathbb C}}

\def\bbr{{\mathbb R}}

\def\bbz{{\mathbb Z}}

\def\tr{\mathop{\rm tr}\nolimits}

\def\frac#1#2{{{#1}\over{#2}}}
\def\tfrac#1#2{{\textstyle{{#1}\over{#2}}}}

\def\half{\tfrac{1}{2}}
\def\third{\tfrac{1}{3}}

\def\sixth{\tfrac{1}{6}}

\begin{document}

\begin{titlepage}

\baselineskip 24pt

\begin{center}

{\Large {\bf Exploring Framed Gauge Theory as Basis for 
Physical Models}}

\vspace{.5cm}

\baselineskip 14pt

{\large CHAN Hong-Mo}\\
h.m.chan\,@\,stfc.ac.uk \\
{\it Rutherford Appleton Laboratory,\\
  Chilton, Didcot, Oxon, OX11 0QX, United Kingdom}\\
\vspace{.2cm}
{\large TSOU Sheung Tsun}\\
tsou\,@\,maths.ox.ac.uk\\
{\it Mathematical Institute, University of Oxford,\\
  24-29 St. Giles', Oxford, OX1 3LB, United Kingdom}

\end{center}

\vspace{.3cm}

\begin{abstract}

It is shown that by introducing as dynamical variables 
in the formulation of gauge theories the frame vectors 
(or vielbeins) in internal symmetry space, in addition to 
the standard gauge boson and matter fermion fields, one 
obtains: (i) for the $su(2) \times u(1)$ symmetry, the
standard electroweak theory with the Higgs field thrown
in as part of the framed gauge theoretical structure,
(ii) for the $su(3) \times su(2) \times u(1)$ symmetry, 
a ``framed standard model'' with, apart from the Higgs
field as before, a global $su(3)$ symmetry to play the
role of fermion generations, plus some other properties
which are shown elsewhere to give to both quarks and 
leptons hierarchical mass and mixing patterns similar 
to those experimentally observed.  Besides, the ``framing'' 
of the standard model as such has brought the particle 
theory closer in structure to the theory of gravity where 
vierbeins have long figured as dynamical variables.  
Although most of the results have already been reported 
before, time and hindsight have allowed their presentation 
in this review to be made more transparent and succint.

\end{abstract}

\end{titlepage}

\clearpage

\section{Introduction}

It can be said that all physical theories we know today are gauge
theories.  The standard model of particle physics is a {\it 
bona fide} gauge theory which has yielded up to the present
an excellent description of all particle physics data, i.e.\  
in effect, we believe, all known physical phenomena besides 
gravity, while gravity itself can be considered as a sort of 
gauge theory as well.  Unfortunately, this rather grandiose 
picture is spoiled by the following facts.  On the one hand, 
in present formulations of the standard model, besides gauge 
principles based on geometry, some extraneous phenomenological 
inputs 
are needed, such as a Higgs scalar to break the 
electroweak symmetry, and 3 generations of fermions together 
with all their peculiar mass and mixing patterns.  On the 
other hand, the interpretation of gravity as a gauge theory 
involves further twists, namely that the gauge group has to 
be soldered to space-time and that the space-time metric is 
to function as dynamical variables.  It would be conceptually 
much more appealing if one can somehow remove from the 
standard model the necessity for those injections from 
experiment, and at the same time to put it with gravity in 
some common gauge theoretical framework.  Even practically, 
it may be rewarding if this can be done, for it would very 
likely help to reduce the large number of parameters needed 
to be fed in from experiment by the standard model, which 
subtracts much from our faith in it as a candidate for a 
fundamental theory.

The purpose of this paper is to review a proposal first made 
some years ago \cite{physcons} and further developed since,
which will be referred to here as the framed gauge theory
(FGT) framework.  This suggests that one includes as dynamical 
variables in the formulation of a gauge theory the frame 
vectors in internal symmetry space, in addition to the usual 
gauge vector boson and matter fermion fields.  These frame 
vectors or vielbeins, one can argue, are an integral part of 
the gauge structure since they have to be there to specify 
the frames to which the gauge transformations refer.  That 
they can be taken as dynamical variables is familiar already 
in the theory of gravity where vierbeins are often taken in 
place of the metric as dynamical variables.  Indeed, by 
introducing now the frame vectors (or ``framons'') also in particle 
physics as dynamical variables, one might have made the first 
tentative step towards its desired rapprochement with gravity, 
for without the framons, there has been so far in the standard 
model no analogue to the metric in gravity.

But of what good is FGT to particle physics itself?  Consider
first just the electroweak sector.  There, experiment tells us
that the gauge symmetry $su(2) \times u(1)$ has to be broken, 
and for that one introduces in the standard electroweak theory
an $su(2)$ doublet of Higgs scalar fields.  But that is exactly 
what a framon in $su(2)$ would look like, for being a frame 
vector, it transforms as the fundamental representation 
of the symmetry, i.e., a doublet in $su(2)$, but under Lorentz 
transformations in space-time, it is a scalar.  In other 
words, by adopting the FGT philosophy, one would obtain as part 
of the gauge structure automatically the Higgs scalar that one 
needs.  There are, of course, two frame vectors in $su(2)$ 
which are orthogonal to each other, but these are neatly
paralleled by the quantities usually denoted by $\phi$ and 
$\phi^c$ in the standard electroweak theory, which are what
appear in the Yukawa term coupled respectively to the up and
down components of the quark or lepton fields.  That being 
the case, one would not be too surprised to find, as will be 
shown later, that in the FGT language the standard electroweak 
theory appears just as the ``minimally framed'' gauge theory 
with gauge symmetry $su(2) \times u(1)$, but with the Higgs 
boson now forming an integral part of the gauge structure.

But what about the strong sector?  If one were to take there
the frame vectors in $su(3)$ colour as dynamical variables as
FGT implies, would it not break the colour symmetry where we
want colour to be confining and exact?  Intriguingly, this
need not be the case.  Return first to the electroweak theory, 
where we recall that in place of the picture usually adopted 
of the local gauge symmetry $su(2) \times u(1)$ being broken
spontaneously by Higgs fields, one could equally have adopted 
instead the picture \cite{tHooft,Bankovici} where the local 
symmetry is confining; what is broken is only a global 
symmetry which one may call the ``dual'' $\widetilde{su}(2) 
\times \tilde{u}(1)$ to the local one above.  Indeed, one may 
even prefer this ``confinement picture'' to 
the usual one of spontaneous breaking of the local symmetry 
as a physically more appealing interpretation of the same 
symmetry-breaking phenomenon.  Applying then this confinement 
picture to colour $su(3)$ above, one concludes correspondingly 
that what FGT implies is only that there is a global symmetry 
$\widetilde{su}(3)$ ``dual'' to the original local colour $su(3)$ 
that is broken, while $su(3)$ colour itself remains confining 
and exact.

The appearance of a new global $su(3)$ symmetry in particle 
theory would in fact be welcome, for the existence of 3 fermion 
generations in nature has already suggested to many such a 
symmetry \cite{horizontal}.  But can the particular 
$\widetilde{su}(3)$ here play 
the desired role of fermion generations?  This symmetry arises in
FGT simply from the fact that frame vectors, by their very nature, 
have to carry two types of indices, one type referring to the
local and the other to a global reference frame.  Recall as
example the vierbeins in gravity, usually denoted as $e^a_\mu$
and labelled by the two types of indices $\mu$ and $a$.  Since
physics should be independent of the choice of reference    
frames, gravity is invariant under Lorentz transformations in
$a$.  For the same reason then, particle physics should also
be invariant under $\widetilde{su}(3)$.  Initially, only 
framons carry this global index, say $\tilde{a} = \tilde{1},
\tilde{2},\tilde{3}$, while all other fundamental, including 
fermionic, fields would carry only the local index, say $a = 
1,2,3$.  Remember, however, that in the confinement picture 
of the electroweak theory \cite{tHooft,Bankovici}, quarks and 
leptons appear not as fundamental fermion fields but as bound 
states, via $su(2)$ confinement, of these with the fundamental 
scalars, i.e.\ in the FGT scenario with the framons, and can 
acquire therefore from the latter the 3-valued global index 
$\tilde{a}$ to play the role of fermion generations.

In other words, it would appear that the FGT framework contains 
in it already the ingredients for providing not only a Higgs 
field necessary for breaking the electroweak symmetry but also 
exactly 3 generations of quarks and leptons as experiments seem
to demand.  But can it really do so in practice?  Encouraged 
by the above observations, let us proceed to construct, as one 
did successfully, it seems, for the electroweak theory, the 
corresponding ``minimally framed'' theory for the gauge symmetry 
$su(3) \times su(2) \times u(1)$.  One obtains then a construct
which one can call the framed standard model (FSM).  As will be
shown later, this is found indeed to give the Higgs field and 3 
generations of quarks and leptons as expected.  Moreover, it is
found that these quarks and leptons appearing as fermion-framon 
bound states via $su(2)$ confinement have mass matrices at tree 
level of the form:
\begin{equation}
m = m_T {\balpha}{\balpha}^\dagger
\label{mfact}
\end{equation}
where the vector ${\balpha}$ is ``universal'', meaning that it 
is independent of the fermion species, i.e.\ whether it is up or 
down in flavour or whether it is a quark or a lepton.  Now, such 
a ``universal'' rank-one mass matrix, giving only one heavy state 
in each species and the unit matrix as the mixing matrix, has 
long been advocated \cite{Fritsch,Harari} as a good zeroth-order 
starting point since it is already not that far from the actual 
situation observed in experiment.

But this is not all.  The FSM is by construction invariant, as 
it ought to be by previous arguments, under both the original 
local gauge symmetry $su(3) \times su(2) \times u(1)$, and its
``dual'', i.e.\ the global symmetry $\widetilde{su}(3) \times 
\widetilde{su}(2) \times \tilde{u}(1)$.  This doubled invariance
places severe restrictions on the form that terms of the action
containing the framon fields can take, in particular on the 
self-interaction potential, say $V[\Phi]$, of the framon field 
$\Phi$.  This $V[\Phi]$, on minimization, will tell us what 
the vacuum will look like, and will allow us further to evaluate 
renormalization effects on the vacuum and hence also on the 
vector ${\balpha}$ from the mass matrix (\ref{mfact}).  And
these renormalization effects are found to give automatically
deviations from the zeroth order approximation above and lead to
a hierarchical mass spectrum for both quarks and leptons, as well 
as mixing matrices qualitatively similar to that experimentally 
observed with CP-violation included.  

In other words, it would appear that the FSM when formulated as 
an FGT is capable of reproducing all those idiosyncrasies of the 
standard model mentioned at the beginning such as the Higgs 
boson and the 3 fermion generations together with their mass and 
mixing patterns as consequences of the gauge principle as wanted.  
In what follows, we shall review the procedure whereby those of 
the above results concerning the structure of the model are 
deduced, i.e.\ all apart from those outlined in the last paragraph 
which are derived from the dynamics.  Although most of the 
reviewed material has appeared in some form or another before, 
e.g. in \cite{prepsm}, time and experience have given it the 
greater clarity and cogency now needed when viewed in the present 
wider context.  The derivation of results outlined in the last 
paragraph is not reviewed because it has been completed only 
recently and is reported in separate papers \cite{dfsm,r2m2} 
to which the reader can be referred.

One turns instead to the other structural question raised at the
beginning, namely whether particle physics when formulated now 
as an FGT can be put with gravity on a common gauge theoretical 
footing.  As already noted, by the introduction of frame vectors 
as dynamical variables in particle physics one has already taken 
a first step towards a possible rapprochement with gravity, for
this will mean that particle theory will now acquire also a
variable metric, though here not in space-time but in internal 
symmetry space.    However,
if one were to borrow the Kaluza-Klein idea that internal space
may be part of a larger space-time compactified to a very
small size, then a metric in space-time and in internal space
may not appear as so very different concepts.  Indeed, building 
on this in the last section, one is led to some interesting 
speculations on how the two sides, i.e.\ particle physics on the
one hand and gravity on the other, may perhaps be brought closer
together as just two different parts of the same overarching 
``framed'' gauge theoretical framework. 

Taken all together, these considerations, to be expanded below,
would seem to suggest that the FGT framework could perhaps be 
taken with some credibility, or at least considered worth exploring, 
as a viable basis for the physics we know today.

\section{Frames and Minimal Frames}

First, we need to make precise what is meant above by framing.
Apart from gravity, the framing of which in terms of vierbeins
is already familiar, what interests us here is the framing of
the standard model of particle physics, i.e.\ the gauge theory
with gauge symmetry $su(3) \times su(2) \times u(1)$.  Before
starting on this, however, let us first consider each of its 3
component symmetries, namely $su(3)$, $su(2)$ and $u(1)$.

In the notation here adopted, lower case letters as in $su(N)$ 
denote the algebra but capital letters as in $SU(N)$ the group.  
The same algebra, of course, may correspond to different groups; 
thus, for example, both $SU(2)$ and $SO(3)$ have $su(2)$ as their 
algebra, but $SU(2)$ double covers $SO(3)$.  What specifies which 
group one is dealing with in a given theory is the representations 
which appear in the theory.  For instance, only the $SU(2)$
theory, not $SO(3)$, has doublets as representations.  Since 
the theories we are interested in all contain fields in the 
fundamental representation of $su(N)$, it is with those with 
gauge groups $SU(N)$ that we shall be concerned. 

As for the vierbeins $e_\mu^a$ in gravity, frame vectors in an 
$SU(N)$ theory can be taken as the column vectors of the matrix 
relating the local frame to the global reference frame, say:
\begin{equation}
\Phi = (\phi_a^{\tilde{a}}),
\label{Phi}
\end{equation}
where the row index $a$ refers to the local frame and the column
index $\tilde{a}$ to the global reference frame.  Since both the 
local and global frames here are unitary, the matrix $\Phi$ is 
itself an element of $SU(N)$.  Local $SU(N)$ transformations 
on $\Phi$ act from the left while global $\widetilde{SU}(N)$ 
transformations act from the right.

The frame vectors $\bphi^{\tilde{a}}$ in $su(N)$ space labelled 
by index $\tilde{a}$ transform as fundamental representations 
under the local $su(N)$, but are scalars under space-time 
Lorentz transformations, and they satisfy originally, of course, 
the following conditions:
\begin{itemize}
\item[(a)] They have unit length, 
$|{\bphi}^{\tilde{a}}| = 1$;
\item[(b)] They are mutually orthogonal, ${\bphi}^{\tilde{a}}
     \cdot {\bphi}^{\tilde{b}} = 0,\ \tilde{a} \neq \tilde{b}$;
\item[(c)] The determinant is real.
\end{itemize}

But, in adopting them as dynamical variables as proposed, we 
are in effect promoting them into fields.  They should then be 
allowed to take any complex values as ordinary scalar fields do, 
in which case they will not be able to satisfy all 3 conditions 
above.  However, these $\phi_a^{\tilde{a}}$ need not all be taken 
as independent variables, so that some of the conditions can 
still be retained.  We ask then what is the most economical 
arrangement with the smallest number of ``framons'' introduced.  
We shall say then that the resulting theory is ``minimally 
framed''.

Consider first $SU(2)$ as an example.  There are 2 frame vectors
${\bphi}^{\tilde{1}}$ and ${\bphi}^{\tilde{2}}$ which
satisfy the 3 conditions (a), (b) and (c).  To allow them to 
take any complex values when promoted to framon fields as 
stipulated, we must of course relax the condition (a), but we 
can still keep the other 2 conditions.  Explicitly, since the 
framons are not required to be independent, we can write:
\begin{equation}
\phi^{\tilde{2}}_r =- \epsilon_{rs} (\phi^{\tilde{1}}_s)^*,
\label{su2ortho}
\end{equation}
which will keep the 2 framons orthogonal and of equal length
while keeping also the determinant real, but will still allow both 
to take all complex values.  One needs to introduce then as 
framons in the minimally framed theory only 1 complex vector 
or 4 real scalars.  Alternatively, one can think of this as a 
problem of embedding $SU(2)$ in $\bbr^n$, where it is well known 
that the minimal embedding is as the unit sphere in $\bbr^4$.

Consider next $SU(3)$, where there are 3 frame vectors. 
To promote these into fields
we must again relax condition (a),
but we can no longer retain the condition (b) that the framons 
remain mutually orthogonal as we did for $SU(2)$.  Indeed, if
we write down here the parallel to (\ref{su2ortho}), thus:
\begin{equation}
\phi^{\tilde{3}}_r = \epsilon_{rst} (\phi^{\tilde{1}}_s)^*
   (\phi^{\tilde{2}}_t)^*,
\label{su3ortho}
\end{equation}
we find that this will give the 3 framons enforcedly different 
physical dimensions, which we cannot accept.  We are left then
with only condition (c), namely that the determinant is real,
which still allows the different framons to have the same
physical dimension since the determinant, though complicated, 
is multilinear in all its elements.  The same conclusion holds 
for any $SU(N)$, $N > 2$.

One concludes therefore that for the ``minimally framed theory''
one needs to introduce only 4 real scalar fields as framons for 
the $SU(2)$ theory, but for $SU(N), N > 2$ theory one needs 
in general $2 N^2 - 1$ real scalar fields, e.g. 17 for $SU(3)$.

This leaves now only the $u(1)$ factor still to be considered. 
Here orientation means just a phase, and relative orientation
just a phase difference.  Hence, the analogue of $\Phi$ above 
for the $su(N)$ factors is here a phase factor of the form:
\begin{equation}
\Phi = \exp i g_1 (\alpha - \tilde{\alpha}),
\label{Phiu1}
\end{equation}
with $\alpha$ $x$-dependent, transforming under the local $u(1)$ 
but $\tilde{\alpha}$ $x$-independent, transforming under the 
global $\tilde{u}(1)$.  The framon field is then a complex
scalar field with its phase as in (\ref{Phiu1}) above.  

Having now specified what framing means for each of the theories 
with the symmetries $su(2), su(2)$ and $u(1)$, we are ready to 
tackle the physical theories with the product symmetry.

\section{The Electroweak Theory}

As a warming-up exercise before we proceed to the full standard 
model, let us first consider the electroweak theory, which is 
a gauge theory with the gauge symmetry $su(2) \times u(1)$.  As 
it is usually formulated, the standard gauge principles have to 
be supplemented by the introduction of the Higgs scalar to break 
the $su(2)$ symmetry as required by experiment.  Here we wish 
to approach the problem anew from the viewpoint of framed gauge 
theory (FGT), and try to show that the same electroweak theory 
will emerge as the minimally framed gauge theory for the same 
gauge symmetry, but now with the Higgs scalar thrown in as an 
integral part of the framed gauge theoretical framework. 

In the same spirit then as what was done above in section 2 for 
the $su(N)$ symmetries, we ask first what scalar framon fields 
are to be introduced for the theory to be ``framed''.  The frame 
vectors, and hence also the framons to which they are promoted, 
are to be representations of the symmetry $su(2) \times u(1)$, 
and there are two ways of building representations of a product 
from those of its factors, namely as the product or as the sum.  
Suppose we appeal again to economy, or ``minimality'' for whatever 
it is worth, and ask which gives the smaller number of scalar 
fields, we would favour the product, $2 \times 1$ being smaller 
than $2 + 1$.  Hence, one would introduce as framon fields two
$su(2)$ doublets: $\bphi^{\tilde{r}}$ labelled by the 
global index $\tilde{r} = \tilde{1}, \tilde{2}$, each forming 
also a representation of the local $u(1)$ symmetry, or in other 
words, each carrying also a $u(1)$ (hyper-)charge.

What $u(1)$ charges should they carry?  To answer this, again 
a question of representations, one would need first, as in the
$su(N)$ symmetries above, to specify the gauge group.  There 
are 3 groups all having $su(2) \times u(1)$ as its algebra, namely 
$SU(2) \times U(1), SO(3) \times U(1)$ and $U(2) = (SU(2) \times 
U(1))/\bbz_2$.  By examining the fields present in the standard 
electroweak theory, noting for example that $su(2)$ doublets 
exist with half-integral $u(1)$ (hyper-)charges, one comes to 
the conclusion \cite{ourbook} that $U(2)$ is the gauge group one 
needs.  From this, it follows that the framon fields ${\bphi}
^{\tilde{r}}$ above should also carry half-integral $u(1)$ charges.

Moreover, from the last section, one has learned that for $su(2)$, 
one can further economize on the number of framons introduced by 
insisting that $\phi_r^{\tilde{r}}$ satisfy the orthogonality 
condition (\ref{su2ortho}), leaving then only one doublet independent, 
i.e.\ in total just 2 complex or 4 real scalar fields as variables.
This means also that whatever $u(1)$-charge $\bphi^{\tilde{1}}$
carries, then $\bphi^{\tilde{2}}$ would carry the opposite.  Note
however that $\bphi^{\tilde{1}}$ and $\bphi^{\tilde{2}}$ need 
not themselves be eigenstates of the $u(1)$-charge, which can be 
chosen instead as any two mutually orthogonal linear combinations.  
These we can specify by introducing a vector, say ${\bgamma} = 
(\gamma^{\tilde{r}})$ in $\widetilde{su}(2)$ space, such that 
the following vectors, now back in $su(2)$ space, are eigenstates 
of the $u(1)$-charge with the (hyper-)charges shown:
\begin{eqnarray}
\bphi^{(+)} & = & \sum_{\tilde{r}} \gamma^{\tilde{r}} 
   \bphi^{\tilde{r}};\  y = g_1/2, \nonumber \\ 
\bphi^{(-)} & = & \sum_{\tilde{r}} \gamma'^{\tilde{r}}
   {\bphi}^{\tilde{r}};\  y = -g_1/2,
\label{bphipm}
\end{eqnarray}
where $\bgamma'$ is the vector orthogonal to ${\bgamma}$ in 
$\widetilde{su}(2)$ space.  This notation is useful later when
invariance under $\widetilde{su}(2)$ is considered.  Otherwise, 
it is convenient to choose a frame in $\widetilde{su}(2)$ space 
such that ${\bgamma}$ points in the $(1, 0)$ direction so that 
$\bphi^{\tilde{1}}$ coincides with $\bphi^{(+)}$ as is often 
done in the literature.  Notice that $\bphi^{(+)}$ has exactly 
the same $su(2)$ representation and $u(1)$ charges as the Higgs 
field in the standard electroweak theory.

Next, we turn to the action for the framed theory which we want 
to be invariant not only under the local gauge symmetry $su(2) 
\times u(1)$ we started with, but also under the global symmetry 
$\widetilde{su}(2) \times \tilde{u}(1)$.  Since only the framon 
fields carry the global indices, we need consider here only those 
new terms of the action which contain the framons, as the others, 
such as the gauge field action or the kinetic energy term of the 
fermions, will be the same as in the standard electroweak theory.

Of the new terms containing framons, consider first the potential
term of framon self-interaction which one can take, as usual, to 
be a polynomial of even powers of the framon fields, but only up 
to and including quartic terms for renormalizability.  To ensure
invariance under the double symmetry $su(2) \times u(1) \times
\widetilde{su}(2) \times \tilde{u}(1)$, we saturate all indices
in all possible ways, and end up with the following general form:
\begin{equation}
V[\Phi] = - \mu \sum_{r,\tilde{r}} (\phi_r^{\tilde{r}})^* \phi_r
   ^{\tilde{r}} + \lambda \left(\sum_{r,\tilde{r}} (\phi_r^{\tilde{r}})^* 
   \phi_r^{\tilde{r}}\right)^2 + \kappa \sum_{r,s,\tilde{r},\tilde{s}}
   (\phi_r^{\tilde{r}})^* \phi_r^{\tilde{s}} (\phi_s^{\tilde{s}})^*
   \phi_s^{\tilde{r}},
\label{VPhiEW}
\end{equation}
which can be written more succinctly in terms of the matrix $\Phi$
introduced in (\ref{Phi}) above as:
\begin{equation}
V[\Phi] = - \mu \tr(\Phi^{\dagger} \Phi) 
   + \lambda \left(\tr(\Phi^{\dagger} \Phi)\right)^2
   + \kappa \tr(\Phi^{\dagger} \Phi \Phi^{\dagger} \Phi).
\label{VPhiEWa}
\end{equation}
As written, this depends on both the vectors $\bphi^{\tilde{1}}$ 
and $\bphi^{\tilde{2}}$, of which, however, only one is independent.  
Eliminating, say, $\bphi^{\tilde{2}}$ in terms of $\bphi^{\tilde{1}}$ 
using the orthogonality condition (\ref{su2ortho}), one is left with 
the usual Mexican hat potential of the standard electroweak theory 
with $\bphi^{\tilde{1}}$ identified with the Higgs field there.  That 
this is so can most easily seen by again rewriting (\ref{VPhiEW})
in terms of the vectors $\bphi^{\tilde{r}} = (\phi_r^{\tilde{r}})$,
thus:
\begin{equation}
V[\Phi] = - \mu \sum_{\tilde{r}} |\bphi^{\tilde{r}}|^2
           + \lambda \left(\sum_{\tilde{r}} |\bphi^{\tilde{r}}|^2\right)^2
           + \kappa \sum_{\tilde{r},\tilde{s}} 
             |(\bphi^{\tilde{r}})^{\dagger}\cdot\bphi^{\tilde{s}}|^2.
\label{VPhiEWb}
\end{equation}
Since what the condition (\ref{su2ortho}) says is that the two vectors 
$\bphi^{\tilde{1}}$ and $\bphi^{\tilde{2}}$ are mutually orthogonal
and have the same length, it follows immediately that (\ref{VPhiEWa}) 
is reduced to:
\begin{equation}
V[\Phi] = - 2 \mu |\bphi|^2 + (4 \lambda + 2 \kappa) |\bphi|^4
\label{VPhiEWbb}
\end{equation}
as claimed.

Secondly, consider the kinetic energy term of the framons, which
we can write most succinctly in terms of the matrix $\Phi$ as:
\begin{equation}
\tr((D_\mu \Phi)^{\dagger} D_\mu \Phi),
\label{KEofPhi}
\end{equation}
with
\begin{equation}
D_\mu \Phi = \partial_\mu \Phi - ig_2 B_\mu \Phi 
   - \half i g_1 A_\mu \Phi \Gamma,
\label{DmuPhi}
\end{equation}
where $\Gamma$ is the matrix:
\begin{equation}
\Gamma = (\bgamma, -\bgamma') = \left( \begin{array}{cc}
   \gamma^{\tilde{1}} & - \gamma'^{\tilde{1}} \\ \gamma^{\tilde{2}} 
& - \gamma'^{\tilde{2}} 
   \end{array} \right),
\label{Gamma}
\end{equation}
so that by construction:
\begin{equation}
\Phi \Gamma = (\bphi^{(+)}, - \bphi^{(-)}) 
   = \left( \begin{array}{cc} \phi_1^{(+)} & - \phi_1^{(-)} \\
   \phi_2^{(+)} & - \phi_2^{(-)} \end{array} \right),
\label{PhiGamma}
\end{equation}
thus giving the correct $u(1)$ charges, $\pm g_1/2$ respectively, 
to the two vectors $\bphi^{(\pm)}$.  The term (\ref{KEofPhi}) is
explicitly invariant both under all local $su(2) \times u(1)$
and under all global $\widetilde{su}(2) \times \tilde{u}(1)$
transformations, as required.

To show that the term (\ref{KEofPhi}) above under the condition 
(\ref{su2ortho}) is in fact the same as the corresponding term in the
standard electroweak theory, all we need is to choose $\bgamma$ to
be real and point in the up direction, making thus $\bphi^{(+)}$
the same as $\bphi^{\tilde{1}}$.  A direct calculation then shows
that, because of (\ref{su2ortho}), the two terms coming respectively
from $\bphi^{(+)}$ and $\bphi^{(-)}$ are in fact identical and add
up for $\bphi^{(+)} = \bphi$ to just:
\begin{equation}
2 (D_\mu \bphi)^{\dagger} D_\mu \bphi,
\label{KEofphiEW}
\end{equation}
with
\begin{equation}
D_\mu = \partial_\mu - i g_2 B_\mu - \half i g_1 A_\mu,
\label{DmuphiEW}
\end{equation}
i.e.\ of the form familiar in the standard electroweak theory.

Lastly, we need to consider the Yukawa term coupling the framon
to the fermion fields, which we can write as usual as:
\begin{equation} 
Y \bar{\psi}^r \phi_r^{(-)} \half (1 + \gamma_5) \psi
    + Y' \bar{\psi}^r \phi_r^{(+)} \half (1 + \gamma_5) \psi'
    + {\rm h.c.}
\label{Yukawaw}
\end{equation}
with only the proviso that $\bphi^{(\pm)} = (\phi_r^{(\pm)})$ is now 
to be taken in general as (\ref{bphipm}) above which exhibits its
required invariance also under $\widetilde{su}(2)$. 

One sees therefore that in the framed gauge theory language, the
minimally framed theory with gauge symmetry $su(2) \times u(1)$
is indeed just the standard electroweak theory as claimed, but
now with the framon field playing the role of the standard Higgs
scalar already built in as part of the gauge structure.  For the 
electroweak theory itself, this is merely a formal gain, but as 
we shall see, when the same considerations are applied to the 
standard model, we shall arrive at more substantial results.

Before we proceed further, however, let us first recall some old
results of 't~Hooft \cite{tHooft} and of Banks and Rabinovici 
\cite{Bankovici} which will be of use later.  The theory given
by the action just derived for the symmetry $su(2) \times u(1)$
is usually interpreted as one in which the local symmetry is 
spontaneously broken.  But, as these authors have shown, it may
equally be interpreted as a theory in which the $su(2)$ symmetry
confines and remains exact; what is being broken is only a global
symmetry associated with it which can be identified with what is
denoted by $\widetilde{su}(2)$ above.  This global symmetry is
broken explicitly by the choice of the vector ${ \bgamma}$ in
$\widetilde{su}(2)$ space which specifies the eigenstates with a
definite $u(1)$ charge $\pm g_1/2$.  So it can be said, as did
't~Hooft, that it is electromagnetism which breaks that global
symmetry.  In this interpretation, or ``confinement picture'' as it
will be called in what follows, only $su(2)$ neutral states can
appear as physical particles, hence neither the $su(2)$ doublet
scalar and fermion fields, nor the $su(2)$ triplet gauge boson
fields, can appear as free particles.  
The physical states known to us are all bound states 
formed via $su(2)$ confinement out of the fundamental scalar and 
fermion fields.  Thus, for example, the Higgs scalar $h$ and the
vector bosons $W$-$Z$ appear as respectively the ``$s$-wave''
and ``$p$-wave'' bound states of a framon-antiframon pair:
\begin{equation}
tr(\Phi^{\dagger} \Phi) \sim F^2 + 2Fh + \ldots,
\label{hbound}
\end{equation}
\begin{equation}
\Phi^{\dagger}(\partial_\mu - i g_2 B_\mu) \Phi 
   \sim i g_2 \tilde{B}_\mu,
\label{WZbound}
\end{equation}
where $F$ represents the vacuum expectation value of the framon
field, while the leptons and quarks appear as bound states of a 
framon with a fundamental fermion, respectively:
\begin{equation}
\Phi^{\dagger} \bpsi \sim \bchi,
\label{lbound}
\end{equation}
\begin{equation}
\Phi^{\dagger} \bpsi_a \sim \bchi_a,
\label{qbound}
\end{equation}
with $a$ in the latter the colour index.  Notice that although 
both these leptons and quarks are by construction singlets in
$su(2)$, they are both  doublets in $\widetilde{su}(2)$, having 
each acquired from its framon constituent an $\widetilde{su}(2)$ 
index, and it is the latter global symmetry which now plays the 
role of the up-down flavour in the confinement picture.  Hence, 
this symmetry being  broken by electromagnetism as explained 
above, it will give different masses for up and down flavoured 
states.  Although in the way the theory is at present applied, 
the confinement picture is mathematically equivalent \cite{tHooft} 
to the usual spontaneous breaking picture as interpretations of 
the same electroweak theory, some may find one more physically 
appealing than the other.  In what follows for the standard model, 
we shall adopt the confinement picture as the more convenient for 
our purpose.

\section{The Framed Standard Model}

The standard model is a gauge theory with the gauge symmetry 
$su(3) \times su(2) \times u(1)$.  Our first question, as with
the electroweak theory before, is what scalar framon fields are
to be introduced so as to make the theory ``framed''.  Framons
are to be representations of the local gauge symmetry.  Again,
for a product symmetry, there are two choices, either the sum
or the product representation, and this applies to the product
between any pair.  If we appeal as before to economy for the
smallest number of real scalar framon fields we have to add,
we shall end up again with the product representation for both
$su(3) \times u(1)$ and $su(2) \times u(1)$, but the sum
representation for $su(3) \times su(2)$ since $3 + 2 < 3 \times 2$. 
In other words, we shall end up with the overall representation 
$(su(3) + su(2)) \times u(1)$, i.e.\ $({\bf 3}+{\bf 2}) \times {\bf
  1}$.   

Next, when taken together, as in section 2, the framons should 
form a matrix transforming from the left as a representation
of the local symmetry, here $su(3) \times su(2) \times u(1)$, 
but from the right as a representation of its ``dual'', i.e.\  
the global symmetry $\widetilde{su}(3) \times \widetilde{su}(2)
 \times \tilde{u}(1)$.  So again the question arises as to 
which representation of the global symmetry it should belong.  
Here, the criterion of ``minimal framing'' is no guide, since the 
symmetry being global, whatever choice of representation 
will lead to the same number of scalar framon fields.  If one 
were to choose $ ({\bf \tilde{3}} + {\bf \tilde{2}}) \times 
{\bf \tilde{1}}$, identical to the choice above for the local 
symmetry, the theory would just break up into two separate 
theories, i.e.\ the electroweak theory plus chromodynamics 
disjoint from each other, which is neither an interesting nor a 
realistic situation.  One opts instead therefore, by invoking 
the anthropic principle perhaps, for the more interesting 
all-product representation ${\bf \tilde{3}} \times {\bf \tilde{2}} 
\times {\bf \tilde{1}}$.

As the result of these considerations, one is then led to 
introduce, for the minimally framed gauge theory with gauge 
symmetry $su(3) \times su(2) \times u(1)$, the following two 
types of framons; first the ``weak'' framons which transform as 
doublets under local $su(2)$ but are invariant under local 
$su(3)$:
\begin{equation}
\phi_r^{\tilde{r} \tilde{a}} = \alpha^{\tilde{a}} \phi_r^{\tilde{r}},
    \ \  \tilde{a} = \tilde{1}, \tilde{2},\tilde{3}, 
    \ \  r = 1, 2, \ \  \tilde{r} = \tilde{1}, \tilde{2}, 
    \ \ \ \  y = \pm 1/2, \ \  \tilde{y} = \mp 1/2,
\label{wframon}
\end{equation}
and secondly the ``strong'' framons which transform as triplets 
under local $su(3)$ but are invariant under local $su(2)$:
\begin{equation}
\phi_a^{\tilde{r} \tilde{a}} = \beta^{\tilde{r}} \phi_a^{\tilde{a}},
    \ \ a = 1, 2, 3, \ \   \tilde{a} = 1, 2, 3, 
    \ \ \tilde{r} = 1, 2,
    \ \ \ \  y = - 1/3, \ \  \tilde{y} = 1/3. 
\label{sframon}
\end{equation}
The $\phi$'s in (\ref{wframon}) and (\ref{sframon}) are local,
i.e.\ $x$-dependent, quantities, while the $\alpha$'s  and $\beta$'s 
are global, i.e.\ $x$-independent, with $\phi_r^{\tilde{r}},
\beta^{\tilde{r}}, \tilde{r} = \tilde{1}, \tilde{2}$ transforming 
as doublets under $\widetilde{su}(2)$ and $\phi_a^{\tilde{a}},
\alpha^{\tilde{a}}, \tilde{a} = \tilde{1}, \tilde{2}, \tilde{3}$, 
transforming as triplets under $\widetilde{su}(3)$.  Hence both 
types of framons, by construction, transform as the product 
representation ${\bf \tilde{3}} \times {\bf \tilde{2}}$ under the 
global symmetry $\widetilde{su}(3) \times \widetilde{su}(2)$, as 
stipulated, but we have yet to justify the assignments above in 
(\ref{wframon}) and (\ref{sframon}) for their $u(1)$ charge $y$ 
and $\tilde{u}(1)$ charge $\tilde{y}$.

As in the electroweak theory before, to assign them appropriate 
$u(1)$ charges, one needs first to specify the gauge group.  There 
are again several groups corresponding to the algebra $su(3) \times 
su(2) \times u(1)$, but by examining the representations of all the 
fields appearing in the standard model, one concludes \cite{ourbook}
that the gauge group is that group obtained by identifying in the 
covering group $SU(3) \times SU(2) \times U(1)$ the following 
sextets of elements:
\begin{eqnarray}
\lefteqn{
(c, f, y) = (\omega c, f, z^4 y) = (\omega^2 c, f, z^2 y)}\nonumber \\
&& {} 
= (c, - f, z^3 y)= (\omega c, -f, z y) = (\omega^2 c, -f, z^5 y)\;,
\label{U23}
\end{eqnarray}
where $c$, $f$, $y$, are elements in respectively in the groups $SU(3)$, 
$SU(2)$, and $U(1)$, and
\begin{equation}
z = \exp i \pi/3
\label{z}
\end{equation}
with $\omega$ being the cube root of unity, which group one can 
call here $U(3,2,1)$.  With $U(3,2,1)$ as gauge group, the allowed
representations are \cite{ourbook}: 
\begin{eqnarray}
&(1, 1); \ \ \ & y = 0 + n, \nonumber \\
&(1, 2); \ \ \ & y = \half + n, \nonumber \\ 
&(3, 1); \ \ \ & y = -\third + n, \nonumber \\
&(3, 2); \ \ \ & y = \sixth +n,
\label{yadmit}
\end{eqnarray}
where the first number inside the brackets denotes the dimension 
of the representation of $su(3)$ and the second number that of 
$su(2)$, and $n$ can be any integer, positive or negative. It then 
follows that one has the $u(1)$ and $\tilde{u}(1)$ charges for the
framons as given in (\ref{wframon}) and (\ref{sframon}), where we 
have kept for simplicity only those charges with the smallest 
allowed absolute values. 
 
These are then the framon fields that are to be introduced for the
minmally framed theory with gauge symmetry $su(3) \times su(2) 
\times u(1)$.  They are not all independent, but according to the 
analysis in section 2, the weak framons $\phi_r^{\tilde{r}}$ are 
to satisfy the condition (\ref{su2ortho}) while the strong framons 
$\phi_a^{\tilde{a}}$ the condition that their determinant is real,
leaving thus altogether 21 independent real scalar fields in the 
theory.

A distinguishing feature of these framon scalars, of course,
is that they carry, in addition to the local indices $r$ and
$a$ which they share with the standard gauge boson and matter 
fermion fields, the global indices $\tilde{r}$ and $\tilde{a}$.
One needs to ask then what physical significance these global
indices possess.  In other words, one would wish to know which 
physical particles carry these global indices as quantum numbers. 
Recalling now the confinement picture of 't~Hooft and of Banks 
and Rabinovici for the interpretation of symmetry-breaking in the 
electroweak theory, one sees that none of the framon fields as 
listed can manifest themselves as actual particles, since they 
all carry $su(2)$ and $su(3)$ indices and have to be confined.  
They have to form bound or confined states either with each other 
or with other fields carrying these indices, and only those states 
thus formed which are neutral under both $su(2)$ and $su(3)$ can 
appear as actual particles.  Those confined by $su(2)$ would appear 
to us now, at the present level of our experimental capability, as 
elementary, but those confined by colour $su(3)$ alone are hadrons, 
which we have learnt already by experiment to penetrate and resolve 
into their coloured constituents.  At this level then, which we may 
call the standard model scenario, we need take account only of the 
deeper confinement by $su(2)$.  Let us then ask in this standard 
model scenario, what $su(2)$ neutral bound states will appear which 
have weak framons as constituents and which will carry, by virtue 
of the weak framon(s) they contain, these global indices as quantum 
numbers.  The examination of such particles would reveal to us the 
sought-for physical significance, if any, that these global indices 
possess.

First, the weak framons can form $su(2)$-neutral bound state 
with their own conjugates via $su(2)$ confinement, saturating
thereby the $su(2)$ indices $r$ which appear in (\ref{wframon}) 
as follows:
\begin{equation}
h_W = \sum_{\tilde{a} r \tilde{r}} \alpha^*_{\tilde{a}} 
   (\phi_{\tilde{r}}^r)^* \phi_r^{\tilde{r}} \alpha^{\tilde{a}},
\label{hWFSM}
\end{equation}
or else via the $su(2)$ gauge bosons $B_\mu$ as follows:
\begin{equation}
W_\mu =  \sum_{\tilde{a} r s \tilde{r}} \alpha^*_{\tilde{a}} 
   (\phi_{\tilde{r}}^r)^* (\partial_\mu - ig_2 B_\mu^{rs}) 
   \phi_s^{\tilde{r}} \alpha^{\tilde{a}}.
\label{WmuFSM}
\end{equation}
These are the exact parallels of (\ref{hbound}) and (\ref{WZbound})
in section 3 for the electroweak theory, called by 't~Hooft there 
respectively the $s$-wave and $p$-wave bound states, of the 
framon-antiframon pair.  We notice that the extra global indices 
$\tilde{a}$ are here summed over and do not in the end figure, and 
one has not yet learned anything new about them.

Secondly, again as in the electroweak theory, in parallel 
to (\ref{lbound}) and (\ref{qbound}) in the preceding section, the 
weak framons of (\ref{wframon}) can also form with the fundamental 
fermion fields $\bpsi = (\psi_r)$ and $\bpsi_a = (\psi_{ra})$ the
following bound states:
\begin{equation}
\chi^{\tilde{r} \tilde{a}} 
= \sum_{r} \alpha^{* \tilde{a}} \phi_r^{*\tilde{r}} \psi_r,
\label{LFSM}
\end{equation}
and 
\begin{equation}
\chi_a^{\tilde{r} \tilde{a}}
= \sum_{r} \alpha^{* \tilde{a}} \phi_r^{*\tilde{r}} \psi_{ra},
\label{QFSM}
\end{equation}
which, in the confinement picture, are to be interpreted respectively as  
leptons and quarks, and these now have to carry the same global 
quantum numbers as their weak framon constituents since their
fundamental fermion constituents carry none.  They will carry the 
2-valued global index $\tilde{r}$ which we have already learned 
before to interpret as up-down flavour.  But they will now also 
carry a new 3-valued global index $\tilde{a}$, which can play the 
role of the fermion generation index.  Of course, as to whether
$\tilde{a}$ can actually function for leptons and quarks as the 
generation index is a question which can only be answered by 
a detailed study of its properties, a question on which we shall 
devote much attention below, but that such an index does emerge
automatically from framing seems already quite interesting. 

This is not all.  The weak framons (\ref{wframon}) carry also 
a global $\tilde{u}(1)$ charge $\tilde{y}$.  Because of the 
intrinsic $\tilde{u}(1)$ invariance built into the theory, 
this $\tilde{u}(1)$ charge is necessarily conserved.  So what 
physical significance does it possess?  Again, the $\tilde{y}$ 
charges from the framon-antiframon pair cancel in (\ref{hWFSM}) 
and (\ref{WmuFSM}) giving a zero value for both the Higgs and 
the $W - Z$ bosons, but it does not now cancel in (\ref{LFSM}) 
and (\ref{QFSM}), since only the framon, but not the fermion,
constituents carry this $\tilde{u}(1)$ charge, which thus
takes the value $\tilde{y} = \pm \half$ for both leptons and
quarks.  Recalling now from (\ref{yadmit}) that 
the fundamental fermions fields $\psi_r$ and $\psi_{ra}$ carry 
respectively the $u(1)$ charges $y =- \half, \sixth$, one has for 
leptons in (\ref{LFSM}) and quarks in (\ref{QFSM}) respectively 
$y = -\half \mp \half, \sixth \mp \half$.  Hence it follows that:
\begin{equation}
\tilde{y} = - y + \half(B - L),
\label{BminusL}
\end{equation}
which is then the physical meaning of $\tilde{y}$ that we seek. 
Given that $y$ is itself a conserved quantity from the 
$u(1)$-invariance of the theory, it follows from the conservation 
of the $\tilde{u}(1)$ charge $\tilde{y}$ that the global quantum 
number $B - L$ has also to be conserved.  It would thus seem 
that one has found here  a gauge 
principle \cite{LeeYang} , namely $\tilde{u}(1)$-invariance peculiar to the
framed gauge theoretical framework, from which baryon number 
conservation (in its modern form of $B - L$ conservation) would 
emerge as a consequence.

These conclusions on the global indices are summarized in the
Table \ref{globalsym}, where the entries in the last row are yet
to be discussed.  One sees that even at this stage, the FSM seems 
already to have offered answers to two questions posed at the 
beginning, first, on the origin of the Higgs boson and second, 
tentatively, also on the origin of fermion generations, with 
the unexpected bonus of $B - L$ conservation thrown in.  

\begin{table}
\begin{eqnarray*}
\begin{array}{||c||c|c|c||}  
\hline \hline
{\rm Symmetry} & \widetilde{su}(3) & \widetilde{su}(2) & \tilde{u}(1) \\
\hline \hline
{\rm Index/Charge} & \tilde{a} & \tilde{r} & \tilde{y} \\ \hline
{\rm Interpretation} & {\rm fermion\ generation} & 
{\rm up/down\ flavour} & B-L \\ \hline
{\rm Status} & {\rm Broken\ by}\ \balpha & 
{\rm Broken\ by}\ \bgamma & {\rm Exact} \\ 
       & {\rm (from\ weak\ sector)} & 
{\rm (from\ e.m.\ sector)} &   \\ \hline \hline 
\end{array}
\end{eqnarray*}
\caption{The global symmetries and their physical interpretations}
\label{globalsym}
\end{table} 

As above in the electroweak theory, our next objective would 
be to construct an action for the framed standard model based on
invariance principles.  Framons having been introduced as field 
variables in FSM, the onus is in principle upon us to include in 
the FSM action all terms which can be constructed with the framons
(\ref{wframon}) and (\ref{sframon}), either by themselves or 
together with the other fields occurring in the theory, namely 
the gauge boson and matter fermion fields, so long as the action 
is invariant under both the original local gauge symmetry $su(3) 
\times su(2) \times u(1)$ and its ``dual'', the global symmetry
$\widetilde{su}(3) \times \widetilde{su}(2) \times \tilde{u}(1)$,
conditional only on it being renormalizable.  As above in the 
electroweak theory, these terms are of three types.  First there
will be a term involving only the framons by themselves, which
we shall call the framon potential $V[\Phi]$.  Secondly, there will
be the framon kinetic energy terms involving the gauge bosons via 
the covariant derivatives of the framons.  Lastly, there are the 
Yukawa terms coupling the framons to fermions.  Of course, there 
will also be terms in the action containing no framons at all, but 
these will be the same as in the ``unframed'' standard model.  Of 
the new terms containing the framon fields, we shall construct and
discuss each in turn below.

Consider first then the framon potential $V[\Phi]$ which will tell 
us about, among other things, the FSM vacuum.  The demand for the 
double invariance under both the local $su(3) \times su(2) \times 
u(1)$ and global $\widetilde{su}(3) \times \widetilde{su}(2) 
\times \tilde{u}(1)$ symmetries stringently constrain the sort 
of terms that can be constructed.  Taking all terms up to fourth 
order in the framon fields (\ref{wframon}) and (\ref{sframon}) for 
renormalizability, and contracting all indices in every way to 
ensure invariance, one is led to a potential of the following 
form:
\begin{equation}
V[\Phi] = V_W[\Phi] + V_S[\Phi] + V_{WS}[\Phi]\,,
\label{VPhi}
\end{equation}
where
\begin{eqnarray}
V_W[\Phi] & = & - \mu'_W \sum_{r,\tilde{r},\tilde{a}}
      \phi_r^{\tilde{r} \tilde{a} *} \phi_r^{\tilde{r} \tilde{a}}
    + \lambda'_W \left[ \sum_{r,\tilde{r},\tilde{a}}
      \phi_r^{\tilde{r} \tilde{a} *} \phi_r^{\tilde{r} \tilde{a}} \right]^2
    +  \kappa_{1W} \sum_{r,s,\tilde{r},\tilde{s},\tilde{a},\tilde{b}}
      \phi_r^{\tilde{r} \tilde{a} *} \phi_r^{\tilde{r} \tilde{b}}
      \phi_s^{\tilde{s} \tilde{b} *} \phi_s^{\tilde{s} \tilde{a}}
      \nonumber \\
  & + & \kappa_{2W} \sum_{r,s,\tilde{r},\tilde{s},\tilde{a},\tilde{b}}
      \phi_r^{\tilde{r} \tilde{a} *} \phi_r^{\tilde{s} \tilde{a}}
      \phi_s^{\tilde{s} \tilde{b} *} \phi_s^{\tilde{r} \tilde{b}}
    + \kappa_{3W} \sum_{r,s,\tilde{r},\tilde{s},\tilde{a},\tilde{b}}
      \phi_r^{\tilde{r} \tilde{a} *} \phi_s^{\tilde{r} \tilde{a}}
      \phi_s^{\tilde{s} \tilde{b} *} \phi_r^{\tilde{s} \tilde{b}},
\label{VPhiW}
\end{eqnarray}
involves only the weak framons (\ref{wframon}),
\begin{eqnarray}
V_S[\Phi] & = & - \mu'_S \sum_{a,\tilde{r},\tilde{a}}
     \phi_a^{\tilde{r} \tilde{a} *} \phi_a^{\tilde{r} \tilde{a}}
    + \lambda'_S \left[ \sum_{a,\tilde{r},\tilde{a}}
      \phi_a^{\tilde{r} \tilde{a} *} \phi_a^{\tilde{r} \tilde{a}} \right]^2
    + \kappa_{1S} \sum_{a,b,\tilde{r},\tilde{s},\tilde{a},\tilde{b}}
      \phi_a^{\tilde{r} \tilde{a} *} \phi_a^{\tilde{s} \tilde{a}}
      \phi_b^{\tilde{s} \tilde{b} *} \phi_b^{\tilde{r} \tilde{b}}
       \nonumber \\
  & + & \kappa_{2S} \sum_{a,b,\tilde{r},\tilde{s},\tilde{a},\tilde{b}}
      \phi_a^{\tilde{r} \tilde{a} *} \phi_a^{\tilde{r} \tilde{b}}
      \phi_b^{\tilde{s} \tilde{b} *} \phi_b^{\tilde{s} \tilde{a}}
    + \kappa_{3S} \sum_{a,b,\tilde{r},\tilde{s},\tilde{a},\tilde{b}}
      \phi_a^{\tilde{r} \tilde{a} *} \phi_b^{\tilde{r} \tilde{a}}
      \phi_b^{\tilde{s} \tilde{b} *} \phi_a^{\tilde{s} \tilde{b}},
\label{VPhiS}
\end{eqnarray}
only the strong framons (\ref{sframon}), and
\begin{eqnarray}
\hspace*{-10mm}V_{WS}[\Phi] & = &
     \!\! \nu_{11} \sum_{r,a,\tilde{r},\tilde{s},\tilde{a},\tilde{b}}
      \phi_r^{\tilde{r} \tilde{a} *} \phi_r^{\tilde{r} \tilde{a}}
      \phi_a^{\tilde{s} \tilde{b} *} \phi_a^{\tilde{s} \tilde{b}}
    + \nu_{21} \sum_{r,a,\tilde{r},\tilde{s},\tilde{a},\tilde{b}}
      \phi_r^{\tilde{r} \tilde{a} *} \phi_r^{\tilde{r} \tilde{b}}
      \phi_a^{\tilde{s} \tilde{b} *} \phi_a^{\tilde{s} \tilde{a}}
      \nonumber \\
  & + & \!\!\nu_{12} \sum_{r,a,\tilde{r},\tilde{s},\tilde{a},\tilde{b}}
      \phi_r^{\tilde{r} \tilde{a} *} \phi_r^{\tilde{s} \tilde{a}}
      \phi_a^{\tilde{s} \tilde{b} *} \phi_a^{\tilde{r} \tilde{b}}
    + \nu_{22} \sum_{r,a,\tilde{r},\tilde{s},\tilde{a},\tilde{b}}
      \phi_r^{\tilde{r} \tilde{a} *} \phi_r^{\tilde{s} \tilde{b}}
      \phi_a^{\tilde{s} \tilde{b} *} \phi_a^{\tilde{r} \tilde{a}}\,,
\label{VPhiWS}
\end{eqnarray}
involves both, linking thus the weak to the strong sector. Next,
recalling the fact that the weak framons (\ref{wframon}) are 
subject to the condition (\ref{su2ortho}), one can simplify and 
rewrite the 3 
terms in $V[\Phi]$ in the following forms:
\begin{equation}
V_W[\Phi] = - \mu_W |\bphi|^2 + \lambda_W (|\bphi|^2)^2,
\label{VPhiW2}
\end{equation}
\begin{equation}
V_S[\Phi] = - \mu_S \sum_{a,\tilde{a}} (\phi_a^{\tilde{a}*}\phi_a^{\tilde{a}})
    + \lambda_S \left[ \sum_{a, \tilde{a}} (\phi_a^{\tilde{a}*}
    \phi_a^{\tilde{a}}) \right]^2 + \kappa_S \sum_{a,b,\tilde{a},\tilde{b}}
    (\phi_a^{\tilde{a}*} \phi_a^{\tilde{b}})
    (\phi_b^{\tilde{b}*} \phi_b^{\tilde{a}}),
\label{VPhiS2}
\end{equation}
\begin{equation}
V_{WS}[\Phi] = \nu_1 |\bphi|^2 \sum_{a,\tilde{a}} \phi_a^{\tilde{a} *}
    \phi_a^{\tilde{a}} + \nu_2 |\bphi|^2 \sum_a \left| \sum_{\tilde{a}}
    (\alpha^{\tilde{a} *} \phi_a^{\tilde{a}})\, \right|^2,
\label{VPhiWS2}
\end{equation}
depending altogether on 7 real parameters $\mu_W, \lambda_W, \mu_S,
\lambda_S, \kappa_S, \nu_1$, and $\nu_2$.

The first thing we would wish to know from the framon potential
$V[\Phi]$ is presumably what it would imply for the vacuum in FSM.
Let us first examine the terms $V_W$ and $V_S$, each involving only 
the weak and strong framons by themselves and see what they imply.
The term $V_W$ is the same as the potential in the electroweak 
theory of which little more need be said at this juncture.  To see 
what $V_S$ would imply for the strong vacuum, it is convenient to 
adopt a vector notation for the strong framons (\ref{sframon}) by
rewriting them as vectors in $\widetilde{su}(3)$ space, labelled 
by the local colour index $a$, 
thus:
\begin{equation}
\bphi_a = (\phi_a^{\tilde{a}}),
\label{bfphia}
\end{equation}
in terms of which $V_S[\Phi]$ then reads as:
\begin{eqnarray}
V_S[\Phi] & = & - \mu_S \sum_a |\bphi_a|^2
          + \lambda_S \left( \sum_a |\bphi_a|^2 \right)^2 \nonumber \\
& & + \kappa_S \sum_a \left( |\bphi_a|^2 \right)^2  
    + \kappa_S \sum_{a \neq b} |\bphi_a^*\cdot\bphi_b|^2.
\label{VPhiSvec}
\end{eqnarray}
We are interested in the situation when the 3 parameters in it,
namely $\mu_S, \lambda_S, \kappa_S$, are all positive, in which
case, as in the familiar $V_W$ of the electroweak theory, the 
vacuum values of $|\bphi_a|$ will be in general nonzero and the 
vacuum degenerate.  Of the terms in (\ref{VPhiSvec}), we see that 
only the second $\kappa_S$ term depends on the orientations of the 
vectors $\bphi_a$, the rest depending only on their lengths.  Hence, 
for $\kappa_S > 0$, the minimum for $V_S$ is attained when $\bphi_a$ 
are mutually orthogonal.  Then, minimizing the remaining terms, which 
are symmetric in $a$, with respect to the lengths of $\bphi_a$, we 
deduce that these lengths should have equal, nonzero values.  In 
other words, we would obtain for the vacuum values of $\bphi_a$ an 
orthonormal triad, as frame vectors are normally expected 
to be.  In passing, we note that since $V_S$ is symmetric under 
$\widetilde{su}(3)$ by construction, so should be its degenerate 
vacuum.  The different vacua in the degenerate set here, however, 
differ from one another only in the orientation of the orthonormal 
triad of frame vectors in the $\widetilde{su}(3)$ or ``generation'' 
space, but otherwise look the same, making thus the degeneracy 
unremarkable.

The situation, however, changes dramatically when the term 
$V_{WS}$ is included, which links the strong to the weak sector.  
In the notation introduced in (\ref{bfphia}), $V_{WS}$ appears 
as: 
\begin{equation}
V_{WS}[\Phi] =  \nu_1 |\bphi|^2 \sum_a |\bphi_a|^2
          - \nu_2 |\bphi|^2 \sum_a |(\balpha^* \cdot \bphi_a)|^2.
\label{VPhiWSvec}
\end{equation}
The first term depends on the weak framon field $\bphi$ only through its 
length $|\bphi|$ and so just modifies the value of the parameter 
$\mu_W$ in the weak potential $V_W$ and will not alter the basic
structure of the weak vacuum.  But $V_{WS}$ contains a $\nu_2$
term involving the relative orientation between the the vectors
$\balpha$ and $\bphi_a$, which means that the strong vacuum 
would be distorted from the snug orthonormal arrangement of frame 
vectors it had before by the vector $\balpha$ coming from the 
weak sector.  

How the vector ${\balpha}$ will affect the vacuum values of 
the strong framons $\phi_a^{\tilde{a}}$ is qualitatively easy 
to see. To be specific, let us take $\nu_2 > 0$ and consider 
first the situation when these framons are kept still having the 
same length, thus allowing only their orientations to vary.  Now 
the $\nu_2$ term in (\ref{VPhiWSvec}) is smallest when the framons 
$\bphi_a$ are all aligned with the vector $\balpha$, but this is 
opposed by the second $\kappa_S$ term in (\ref{VPhiSvec}) which, 
to attain its smallest value, would want instead the framons to 
be mutually orthogonal.  Hence the result of minimizing the two 
terms together would be a compromise where the triad is squeezed 
from orthogonality together towards the vector $\balpha$ which, 
by the symmetry of the problem, would be symmetrically placed 
with respect to the triad.  Consider next the opposite situation 
when the framons are kept mutually orthogonal but allowed only 
to change their lengths relative to one another.  We recall then
that it was the first $\kappa_S$ term in (\ref{VPhiSvec}) whose 
minimization gave the result that the 3 lengths should be equal,
but this is now opposed by the $\nu_2$ term in (\ref{VPhiWSvec})
which, to achieve its smallest value, would prefer to have all 
the length attributed to just one of the framons $\bphi_a$ and 
$\balpha$ to be aligned with it.  Minimizing the two terms 
together would thus once more lead to a compromise where the 
$\bphi_a$'s differ in length from one another and the vector 
$\balpha$ is aligned with the longest. From these two examples, 
it is clear then that when the framons are allowed to change 
both their relative orientations and lengths, there will be 
a trade-off between the two extremes.  In other words, the 
minimum of the potential is degenerate, with a varying amount 
of squeeze on the triad compensated by a simultaneous change in 
the relative framon lengths in a prescribed manner, with the 
vector $\balpha$ snuggling up to, but not exactly aligned with,
the longest framon. 

The properties of the FSM vacuum outlined in the two preceding
paragraphs can be confirmed, of course, by minimizing the 
potential $V[\Phi]$.  This has been done and the result is given 
explicitly in \cite{dfsm}, but for the present discussion, only 
the qualitative features described are needed.

In spite of their very different shapes, however, these vacua in 
the degenerate set must nevertheless be equivalent to one another
under $\widetilde{su}(3)$ transformations, just as for the vacua 
of $V_S[\Phi]$ before the linking term $V_{WS}[\Phi]$ was turned
on, since the whole potential $V[\Phi]$ was constructed to be
invariant under these transformations.  And such they are, as
is shown in \cite{dfsm} with the explicit solution, provided of
course that the $\widetilde{su}(3)$ transformations are applied
not only to the framon vectors $\bphi_a$ but also to the vector
$\balpha$, and the results of the transformations are viewed
each in the appropriate gauge.  However, if $\balpha$ is held
fixed while $\widetilde{su}(3)$ transformations are applied to
the framons alone, then they will appear distorted in different
ways from orthonormality, as outlined in the above paragraph.
It can thus be said that the vector $\balpha,$ coming from the 
weak sector, breaks the $\widetilde{su}(3)$ symmetry in the same
sense that a bar magnet placed in vacuum is said to break space
rotation symmetry.  It is in the same spirit too that it was 
electromagnetism which breaks the $\widetilde{su}(2)$ symmetry 
in the electroweak theory \cite{tHooft} via the vector $\bgamma$ 
in the preceding section.  We thus have the intriguing pattern
of symmetry-breaking entered on the last row of Table 1.  The
breaking of $\widetilde{su}(3)$ and $\widetilde{su}(2)$ are the
same in spirit but not in detail.  On the one hand,
the $\widetilde{su}(3)$ breaking, but not the $\widetilde{su}(2)$
breaking, occurs already in the potential.  On the other hand, in the
$\widetilde{su}(2)$ breaking the preferred direction ${\bgamma}$ is
distinguished by a local gauge interaction, namely that of the
electromagnetic $u(1)$, while for the $\widetilde{su}(3)$ breaking 
the preferred direction ${\balpha}$ is not distinguished by a 
parallel $su(2)$ gauge interaction.   
The reason for these differences can be traced to
the fact that one has chosen, on the basis of what one called
``minimal framing'', the sum representation ${\bf {2}}
+ {\bf {3}}$ for $su(3) \times su(2)$ (hence the global
vector ${\balpha}$) but the product representation 
${\bf {2}} \times {\bf {1}}$ for $su(2) \times u(1)$

What is interesting for the moment is that the different strong
vacua in the degenerate set are each attached to a value of the
vector $\balpha$ which is the same vector which gives a 3-valued
``generation'' index to the leptons and quarks in (\ref{LFSM})
and (\ref{QFSM}) above.  It seems therefore that the strong 
vacuum could have a lot to do, via this vector $\balpha$, with 
the properties of fermion generations, such as their mixing and 
mass hierarchy, and we would be interested now in what way this 
$\balpha$ appears in the lepton and quark mass matrices from 
which these properties are ultimately derived.

To answer this, let us next examine the Yukawa terms
coupling the weak framons to the fundamental fermion fields. 
These terms
give rise to the lepton and quark mass matrices.  To 
do so, we have first to specify what fundamental fermion fields
are to be introduced.  Not having yet given a geometric meaning 
to fermion fields as we think we have to the boson fields, i.e.\ 
both to the vector bosons as gauge (connection) fields and to the 
scalar bosons as framon (frame vector) fields, we have to rely 
for selecting our fundamental fermion fields on information 
gathered otherwise.  Restricting ourselves for simplicity to 
only the fundamental representation of each symmetry, we obtain
the following:
\begin{equation}
\psi(1,1), \psi(3,1), \psi(1,2), \psi(2,3)
\label{fundferm}
\end{equation}
where the first argument denotes the dimension of the $su(3)$
and the second that of the $su(2)$ representation, from which
the admissible $u(1)$ charge for each $\psi$ is then specified
by (\ref{yadmit}).  Of these $\psi$'s, however, not all the
left- or right-handed components are needed.  To see which are
the components needed, recall that in the confinement picture
we have adopted, the left-handed quarks and leptons are flavour
doublet bound states via $su(2)$ confinement of the $su(2)$ 
doublet fundamental fermion fields $\psi$ with the weak framon.  
Then from the fact that in the standard model, based on 
phenomenology, one allows only left-handed flavour doublets and
right-handed flavour singlets of quarks and leptons, it can
easily be shown that here only left-handed $su(2)$-doublet and
right-handed $su(2)$-singlet $\psi$'s are allowed, namely:
\begin{equation}
 \psi_L(1,2), \psi_L(2,3), \psi_R(1,1), \psi_R(3,1).
\label{fundfermp}
\end{equation}

Proceeding with these as the fundamental fermion fields, one has 
then for leptons:
\begin{eqnarray}
{\cal A}_{\rm YK}\!\!\! &=& \!\!\!\sum_{[\tilde{a}] [b]} Y^{\rm lepton}_{[b]} 
    \bar{\psi}^r_{[\tilde{a}]} \alpha^{\tilde{a}} \phi_{r}^{(-)} 
    \half (1 + \gamma_5) \psi^{[b]}
    + \sum_{[\tilde{a}] [b]} Y'^{\rm lepton}_{[b]} 
    \bar{\psi}^r_{[\tilde{a}]} \alpha^{\tilde{a}} \phi_{r}^{(+)} 
    \half (1 + \gamma_5) \psi'^{[b]}
    \nonumber \\
&& {} + {\rm h.c.}
\label{wYukawal}
\end{eqnarray}
and for quarks:
\begin{eqnarray}
{\cal A}_{\rm YK}\!\!\! &=&\!\!\! \sum_{[\tilde{a}] [b]} Y^{\rm quark}_{[b]} 
    \bar{\psi}^{ra}_{[\tilde{a}]} \alpha^{\tilde{a}} \phi_{r}^{(-)} 
    \half (1 + \gamma_5) \psi_a^{[b]}
    +  \sum_{[\tilde{a}] [b]} Y'^{\rm quark}_{[b]} 
    \bar{\psi}^{ra}_{[\tilde{a}]} \alpha^{\tilde{a}} \phi_{r}^{(+)} 
    \half (1 + \gamma_5) \psi_a^{'[b]}
    \nonumber \\
&& {} + {\rm h.c.}
\label{wYukawaq}
\end{eqnarray}
which are of the usual form, except for the appearance of the 
global vector $\balpha= (\alpha^{\tilde{a}})$ carried here by the 
weak framon in (\ref{wframon}).  Notice also that in order to 
saturate the $\tilde{a}$ index carried by ${\balpha}$ so as 
to make the whole invariant under $\widetilde{su}(3)$, one has 
introduced 3 identical copies each of the (left-handed) fermion 
fields $\psi_r$ for leptons and $\psi_{ra}$ for quarks, the 
copies being labelled by the dummy index $[\tilde{a}]$.  Under an 
$\widetilde{su}(3)$ transformation, the terms (\ref{wYukawal})
and (\ref{wYukawaq}) will thus remain invariant only if one 
relabels the fermion fields accordingly, but this should not 
change the physics, given the that these fermion fields are 
$\widetilde{su}(3)$ singlets and are identical otherwise.  This 
is the same argument as was used on the (right-handed) fields 
\cite{Weinberg} so as to write any mass matrix in a hermitian 
form independent of $\gamma_5$, as we shall also do immediately
below.

With these Yukawa terms (\ref{wYukawal}) and (\ref{wYukawaq}), 
one obtains, by substituting for the weak framon its vacuum 
expectation value, say $\zeta_W$, the following mass matrix
for both leptons and quarks:
\begin{equation}
m = \zeta_W \left( \begin{array}{c} \alpha^{\tilde{1}} \\ 
    \alpha^{\tilde{2}} \\ \alpha^{\tilde{3}} \end{array} \right) 
    (Y_{[1]}, Y_{[2]}, Y_{[3]}) \half (1 + \gamma_5)
    +\zeta_W \left( \begin{array}{c} Y^*_{[1]} \\ Y^*_{[2]} \\ Y^*_{[3]} 
    \end{array} \right) (\alpha^{\tilde{1}*}, \alpha^{\tilde{2}*}, 
    \alpha^{\tilde{3}*}) \half (1 - \gamma_5).
\label{massmat}
\end{equation}
Then, again, by relabelling appropriately the right-handed fields
as mentioned in the preceding paragraph, one can rewrite the mass 
matrix for both quarks and leptons in the factorized form 
(\ref{mfact}), with:
\begin{equation}
m_T = \zeta_W \rho_T; \ \ \ 
      \rho_T^2 = |Y_{[1]}|^2 + |Y_{[2]}|^2 + |Y_{[3]}|^2. 
\label{mT}
\end{equation} 

As expected, the mass matrices of quarks and leptons do depend 
on the vector ${\balpha}$.  Besides, they are of rank one, and
expressible as a product of ${\balpha}$ with its hermitian 
conjugate as explained.  But the vector ${\balpha}$, 
originating as it does 
as a factor of the weak framon, is of course independent of which 
fermion the framon is bound to, so that in (\ref{mfact}) only 
$m_T$ depends on the fermion type. Now such a ``universal'' rank-one 
mass matrix for fermions has long been advocated \cite{Fritsch,
Harari} as a good starting point or zeroth-order approximation for 
attacking the fermion mass hierarchy and mixing problems, since  
it has only one massive eigenstate, and it gives for the mixing 
matrix the identity matrix, neither of which conclusions is 
a bad approximation to what is experimentally observed. 

Now, starting with an $\widetilde{su}(3)$ ``generation'' symmetry
as it is done here, it is not trivial to end up with some masses
much larger than others, for any obvious breaking of the symmetry 
would lead to a very different mass pattern.  It is therefore 
quite gratifying that the FSM leads automatically to the above
tree-level mass matrix that phenomenologists have long desired.
However, such a tree-level result is of practical value only 
when one knows how to go further to evaluate higher order effects 
so as to explain the nontrivial mixing between up-down flavours
and the nonzero masses of lower generations actually observed in
experiment.  This last seems difficult in whatever scheme, given 
that it would apparently involve breaking both the ``universality'' 
and the ``factorizabilty'' of the tree-level formula (\ref{mfact})
by subsequent radiative corrections.  For example, to obtain 
nontrivial mixing, whether in leptons or in quarks, one would need 
to make $\balpha$ dependent on up-down flavour.  As far as known, 
however, only the electroweak interactions depend on flavour, and 
they seem too weak to give the desired effects.  This was the case
in the standard model when unframed, and remains so even in the
framed standard model; it was seen already that the weak potential
$V[\Phi]$ in (\ref{VPhiW2}) is the same as before and it can easily 
be seen too that the kinetic energy term is also the same as in
(\ref{KEofPhi}) for the electroweak theory, as the extra factor of
${\balpha}$ carried by the weak framon (\ref{wframon}) is
just traced away.  

Fortunately, however, there is a loop-hole in the above line of 
reasoning which allows nontrivial mixing without breaking the 
universality of $\balpha$, namely when the vector ${\balpha}$ 
happens to depend on scale.  As a global parameter appearing in 
the action, there is in principle no reason why $\balpha$  should 
not acquire scale-dependence under renormalization as coupling
constants in general do.  Whether it actually does in FSM will be 
discussed below, but if we suppose that it 
does, then the earlier conclusions on the quark or lepton masses
and mixing matrices deduced from the mass matrix (\ref{mfact}) will 
have to be reassessed, for it has now to be specified at what scale
each quantity, whether mass or state vector, is to be measured.
Consider first as examples the two heaviest states in each flavour,
namely $t$ in up and $b$ in down.  It was already noted that in 
(\ref{mfact}), the coefficient $m_T$ can depend on flavour and can
thus be identified respectively with $m_t$ for up and $m_b$ for 
down, assuming at the moment for simplicity that $m_T$ itself does 
not depend on scale.  But what are the state vectors of $t$ and $b$
in generation space?  In each case, the state vector should be the 
eigenvector of $m$ in (\ref{mfact}) with the single nonzero value,
namely $\balpha$ itself.  But since $\balpha$ depends on scale by
assumption, one has to ask at what scale in each case to evaluate
this $\balpha$.  Suppose we follow the standard convention and
evaluate it at their respective mass scales, we would have the
state vector for $t$ as ${\bf v}_t = \balpha(\mu = m_t)$ and for
$b$ as ${\bf v}_b = \balpha(\mu = m_b)$.  Hence, the state vectors 
${\bf v}_t$ and  ${\bf v}_b$ for $t$ and $b$ will in general point 
in different directions and give for the CKM matrix element $V_{tb} 
= \langle {\bf v}_t|{\bf v}_b \rangle$ a value different from unity. 
In other words, one would conclude that there will be mixing, quite
contrary to our earlier conclusion from (\ref{mfact}) when $\balpha$
was taken as scale-independent.  And notice that this new conclusion
has been obtained without breaking the universality of $\balpha$,
i.e., without making $\balpha$ at any scale different for the two
different flavours.

Next, what about masses for the lower generations; will they become
nonzero also when $\balpha$ depends on scale?  To answer this, we
need consider only one single flavour, say up, for example.  The
state vector for $t$ we have already identified as ${\bf v}_t = 
\balpha(\mu = m_t)$.  The state vectors ${\bf v}_c$ and ${\bf v}_u$
have both to be orthogonal to ${\bf v}_t$ and have themselves to be
mutually orthogonal, the 3 states being by definition independent
quantum states.  This means, of course, that at $\mu = m_t$, the
states ${\bf v}_c$ and ${\bf v}_u$ both have zero eigenvalues for
mass matrix $m$ in (\ref{mfact}).  But these are not to be taken as
the masses for the $c$ and $u$ quarks, for by the usual convention
adopted above, these masses are to be evaluated at the scale of the
masses themselves, i.e.\ respectively at $\mu = m_c$ and $\mu = m_u$.
At these lower values of the scale, however, the scale-dependent 
vector $\balpha$ would be pointing in different directions than at
$\mu = m_t$, i.e., no longer orthogonal to ${\bf v}_c, {\bf v}_u$,
hence giving nonzero solutions to both $m_c$ and $m_u$.  It is as
if by virtue of this scale-dependence of $\balpha$, some of the
mass carried exclusively by this vector has leaked into the lower
states and imbued them with hierarchical but yet nonzero masses as
experiment indicate.

The possibility outlined above of a scale-dependent $\balpha$ in 
(\ref{mfact}) giving rise both to mixing and a hierarchical mass 
spectrum for both quarks and leptons has in fact been studied
phenomenologically for many years and is found so far to be quite
consistent with existing experimental data.  Cast in this context 
as a hypothesis of a rotating rank-one mass matrix (R2M2), it is 
tested phenomenologically in \cite{btfit} and 
reviewed in some detail in a recent paper \cite{r2m2} to which the 
interested reader is referred.  If sustained, these results from
rotation would relieve us from having to break the universality 
of the mass matrix (\ref{mfact}) with respect to flavour, which
would be a gain in that we would then no longer have to look to 
electroweak forces for the origin of mixing.  The effect can now 
arise in principle via rotation as a result of renormalization in 
the  flavour-independent strong sector which is strength-wise 
much more favourable.  

From the viewpoint taken in the analysis of \cite{r2m2} of mass
matrix rotation as a phenomenological hypothesis, it would appear 
that any theory or model which can generate a mass matrix of form 
(\ref{mfact}) with $\balpha$ rotating sufficiently speedily with 
changing scale would have a fair chance of reproducing the existing 
data on mixing and the mass hierarchy.  Our interest is therefore 
turned next on to the question whether and how such a dependence 
on scale, or rotation, of $\balpha$, or of the mass matrix $m$, can 
indeed arise from strong interaction in a theory.  This would be 
a nontrivial requirement because the mass matrix of both leptons 
and quarks appear originally in the Yukawa term of the electroweak 
sector, and it is not obvious that renormalization effects in the 
strong sector would be transmitted there.

For the framed standard model, however, the interesting thing is
that this will be automatic, as follows.  We recall that the FSM
vacuum is degenerate, with the vacuum values of the strong framons
in the different vacua in the degenerate set being distorted from 
orthonormality in various ways by exactly the vector $\balpha$
coming from the electroweak sector and appearing as a factor in
lepton and quark mass matrices.  Hence, if renormalization effects 
in the strong sector changes the vacuum with changing scale, as 
they normally would, so automatically would $\balpha$ change with 
it.  Thus, the only remaining question is whether the vacuum in 
FSM will indeed change with scale as the result of renormalization 
effects in the strong sector.

This question has been answered recently in the affirmative and
is reported in \cite{dfsm}.  As a result, the vector $\balpha$ 
appearing in the quark and lepton mass matrix (\ref{mfact}) will
indeed rotate with changing scale as postulated in \cite{r2m2} and 
shown there to lead to hierarchical masses and mixing.  Further, 
it is found in \cite{dfsm} that this rotating $\balpha$ will have 
fixed points at scale $\mu = \infty$ and it has already been noted 
just after (\ref{mT}) that it is universal.  In other words,
the mass matrix for quarks and leptons which results from FSM is
shown to possess all those properties which have been identified
in \cite{r2m2} as being essential for a successful description of
the mass and mixing patterns for quarks and leptons observed in
experiment, although an explicit fit to data has yet to be done.

One point of interest worth noting is the following.  In the FSM
scenario outlined above, up-down mixing arises from rotation
while the rotation itself is driven by renormalization effects 
in the strong sector.  So it might appear difficult to obtain in
the mixing matrix a CP-violating Kobayashi-Maskawa phase if the
strong sector is itself CP-conserving, as it is generally believed
to be.  What is intriguing, however, is that rotation, as shown 
in \cite{atof2cps,r2m2}, connects the CP-violating KM phase in 
the CKM matrix to the theta-angle coming from topology in the 
so-called strong CP problem.  Thus, as is also shown in these
papers, removing the theta-angle by a chiral transformation so 
as to make the strong interactions CP-invariant, which one can do
in FSM because the mass matrix is of rank one, will automatically
give a CP-violating phase to the CKM matrix which is
naturally of the order 
of the magnitude experimentally observed, while offering at the 
same time a new solution to the age-old strong CP problem. 

It thus seems that just by the simple added device of ``framing'' 
the gauge theory with a gauge symmetry identical to that of the
standard model, one has gone quite some way towards answering 
the questions posed at the beginning about what we called the 
idiosyncrasies of the standard model.  It offers an explanation 
for the origin of not only the Higgs boson but also the three 
fermion generations together with, qualitatively, their peculiar 
mass and mixing patterns including CP-violation.  
In addition, one has gained two bonuses not initially bargained 
for, namely a gauge origin for $B - L$ conservation and a new 
solution of the strong CP problem.  One has yet, of course, to 
ascertain whether the actual mass and mixing parameters observed 
in experiment can indeed be accommodated in the FSM, and that all 
the various new predictions that the FSM is bound to have would 
either be consistent with existing data, or else are testable by 
future experiments.  This will, of course, be a long drawn-out 
process, some parts of which, we hope, will be dealt with in 
forthcoming papers.  For the rest one would have to rely on future 
scrutiny by the community.  But even at this stage, it seems fair 
to say that the framed standard model, and hence by association 
also the framed gauge theory framework, has displayed sufficient 
features of interest to merit further exploration.

Before one leaves the FSM, there is one more point to be noted 
which will be of use later.  Since frame vectors have been promoted
in FSM to be dynamical variables, so also will be the components
of the metric.  Indeed, we recall that even at vacuum, the frame 
vectors are in general distorted from orthonormality, and so the 
metric will not in general be flat.  Since the original local 
gauge symmetry $su(3)$ is supposedly still to be confining and 
exact, this means that it is the metric:
\begin{equation}
g^{\tilde{a} \tilde{b}} 
   = \sum_a (\phi_a^{\tilde{a}})^*\phi_a^{\tilde{b}}
\label{Gup}
\end{equation}
in the $\widetilde{su}(3)$ or generation space that is distorted 
from flatness.  The details of how this metric is distorted, and 
some implications that this will have on phenomenology, can be 
found in \cite{dfsm}.  We note here only the fact that the metric
in $\widetilde{su}(3)$ is nonflat, which will be relevant for 
the discussion in the next section.

\section{Speculations on Relation to Gravity}

Supposing tentatively that framed gauge theory does serve as a
viable basis for the standard model of particle physics more or
less as it is proposed, it would be natural to ask whether the 
idea can be extended also to gravity which governs the large 
scale physics not covered by particle theory. This is not a
priori hopeless, for we know already that the theory of gravity
is framed (indeed, as stated at the beginning, even the concept 
of framing itself is originally borrowed from gravity), and that
it can be considered as a kind of gauge theory.  The question
only is whether the two theories, particle theory and gravity,
can be brought closer together as to be recognizable as but two
branches of a common framework.

At this stage, let us for the moment throw caution to the wind 
and set our imagination loose on the problem.  Let us first list 
down for this purpose some similarities and differences between 
the two.  In the language of this paper, they are both framed
theories, with each a local and a global symmetry.  For the 
particle theory, the local symmetry is $G$, say, and the global
symmetry $\tilde{G}$, i.e.\ explicitly $su(3) \times su(2) \times 
u(1)$, and $\widetilde{su}(3) \times \widetilde{su}(2) \times
\tilde{u}(1)$.  For gravity, the global symmetry operating in 
the indices $a$ of the vierbeins $e^a_\mu$ is the Lorentz group
which we shall call here $H$ (not $\tilde{H}$, the reason for
which apparent switch in notation between tilde and no-tilde
will soon be obvious).  The local symmetry $\tilde{H}$ we shall
take also to be the Lorentz group.  This at first sight seems
quite off the mark, since the whole point in gravity is that the
metric becomes distorted by matter so that Lorentz invariance is
lost.  However, we have already seen in the framed standard model 
of the preceding section how a framed theory originally symmetric
under $\widetilde{su}(3)$ can have its symmetry broken by the 
interaction and settles to a quite different metric.  So may, we 
think, be the case here too.  Accepting this for the moment, we 
arrange the information in the diagram Fig.\ \ref{Plan1}.

\begin{figure}
\centering
\setlength{\unitlength}{1cm}
\begin{picture}(10,14)
\put(1,12){Internal}
\put(8,12){External}
\put(1.5,11){\Large{$\Xi$}}
\put(8.5,11){\Large{$X$}}
\put(1.5,8){\Large{$\widetilde{G}$}}
\put(8.5,8){\Large{$\widetilde{H}$}}
\put(1.1,7){global}
\put(8.2,7){local}
\put(1.1,6.5){broken}
\put(8.2,6.5){broken}
\put(1.5,3.5){\Large$G$}
\put(8.5,3.5){\Large$H$}
\put(1.2,2.5){local}
\put(8.1,2.5){global}
\put(1.2,2){exact}
\put(8.1,2){exact}
\end{picture}
\caption{Plan 1 for before Kaluza-Klein}
\label{Plan1}
\end{figure}

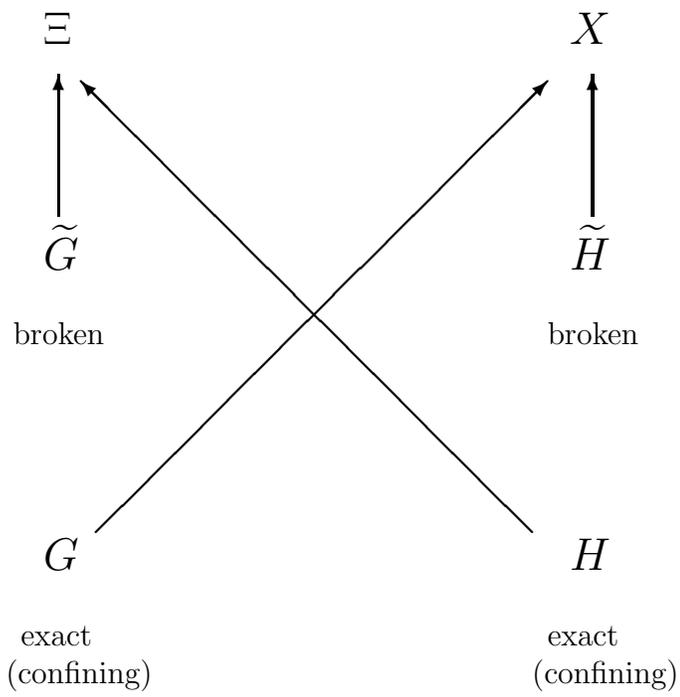
\begin{figure}
\centering
\setlength{\unitlength}{1cm}
\begin{picture}(10,14)
\put(1.5,11){\Large{$\Xi$}}
\put(8.5,11){\Large{$X$}}
\put(1.5,8){\Large{$\widetilde{G}$}}
\put(8.5,8){\Large{$\widetilde{H}$}}
\put(1.1,7){broken}
\put(8.2,7){broken}
\put(1.5,4){\Large$G$}
\put(8.5,4){\Large$H$}
\put(1.2,3){exact}
\put(8.2,3){exact}
\put(1,2.5){(confining)}
\put(8,2.5){(confining)}
\thicklines
\put(1.7,8.7){\vector(0,1){1.9}}
\put(8.8,8.7){\vector(0,1){1.9}}
\put(2.2,4.5){\vector(1,1){6}}
\put(8,4.5){\vector(-1,1){6}}
\end{picture}
\caption{Plan 2 for after Kaluza-Klein}
\label{Plan2}
\end{figure}

Notice that by ``local'' here, we mean that the transformations
of that symmetry can depend on the point $x$ in the ``external''
space $X$, i.e.\ our space-time, and by ``global'' that they do
not, and hence, up to now, they are tacitly taken as constant.
However, suppose we make the Kaluza-Klein assumption that the
``internal'' space $\Xi$ on which the internal symmetries $G$ 
and $\tilde{G}$ operate is compactified and very small in size,
then any quantity depending only on points in $\Xi$ but not on 
points in $X$ would appear to us as effectively constant and be
considered as ``global''.  Accepting this, we are then led to 
the diagram Fig.\ \ref{Plan2}, where an arrow represents the 
assertion that the quantity at the tail of the arrow depends on 
points in the space at its head.  As a result, we see that the 
Kaluza-Klein assumption has made the arrangement now symmetric 
between $\Xi, G, \tilde{G}$ on the one hand and $X, H, \tilde{H}$ 
on the other.

The 2 symmetries $G$ and $\tilde{H}$ remain local in the sense 
that their elements may depend on points in the external space
$X$, but their nature would be quite different.  Whereas for
the $G$-theory, the group elements operate on the internal space
$\Xi$ but depend on points in external space $X$, for the 
$\tilde{H}$-theory, the group elements operate on the same space
$X$, on points of which they can themselves depend.  Explicitly,
suppose we introduce for the $G$-theory $A^{ab}_\mu$ as the gauge 
potential (connection) and from which we construct $F^{ab}_{\mu\nu}$
as the field tensor (curvature), with Latin indices in $\Xi$ but 
Greek indices in $X$.  With no known relationship between the
two types of indices, the simplest invariant that we can construct 
is just $\sum_{ab\mu\nu} F^{ab}_{\mu\nu} F^{ab}_{\mu\nu}$, namely 
the Lagrangian density for the Yang-Mills action.  On the other 
hand, the same considerations will give for the $\tilde{H}$-theory 
the spin connection $\omega^{ab}_\mu$ as the gauge potential and the
Riemann curvature $R^{ab}_{\mu\nu}$ as the field, where both the 
Latin and Greek indices are now in $X$, only referred to different
frames.  A simpler invariant than before can thus be constructed
by contracting Latin with Greek indices by means of the vierbeins
$e^a_\mu$, giving then the scalar curvature $R^{\mu\nu}_{\mu\nu}
= e_a^\mu e_b^\nu R^{ab}_{\mu\nu} = R$, the Lagrangian density
for the Einstein action instead.   

These two ``local'' theories, namely Yang-Mills on the left and 
Einstein on the right, can be so different in nature without 
destroying the symmetry between the two sides of the diagram of 
Fig.\ \ref{Plan2} because they play each a different role there.  
Indeed, according to the diagram, there is in principle also on
the right-hand side a Yang-Mills type theory with $H$ as the 
symmetry group, and on the left-hand side an Einstein type theory
with $\tilde{G}$ as the symmetry group, and both are ``local'' in 
the internal space $\Xi$.  Only, we chose to ignore these theories 
because we took $\Xi$ to be compactified and so small in size as 
to make them inaccessible to us.  In other words, it was the input
of the Kaluza-Klein assumption that $\Xi$ is small while $X$ is 
extended compared to us that gave rise to this difference.  The 
structure of the two sides remain symmetric.

Having noted the symmetry in structure between the two sides of
Fig.\ \ref{Plan2}, let us turn to consider a little the 
possible dynamics.
On the left-hand side, we said in the preceding section for the
particle theory that the $G$ symmetry is exact but the $\tilde{G}$
symmetry is broken, leading thus to a nonflat metric in $\tilde{G}$
-space.  In parallel, we would say on the right-hand side that
the $H$-symmetry is exact but the $\tilde{H}$-symmetry is broken,
leading again to a nonflat metric; this is what we have always 
wanted for gravity, although we do not know yet whether the metric
will be distorted by matter in the correct way.  In the preceding
sections, we also said that the $G$-symmetry is confining so that
only $G$-neutral objects can propagate.  An example is the Higgs 
boson which appeared in this theory as a $G$-neutral bound (confined)
state of a framon-antiframon pair held together by $G$-confinement.
Suppose we now say in parallel on the right-hand side that the
$H$-symmetry is also confining.  Then we would conclude that the
vierbeins $e^a_\mu$ themselves, not being $H$-neutral will have
to be confined, and can propagate only as bound vierbein pairs 
held together by $H$-confinement, namely $\sum_a e^a_\mu e^a_\nu
= g_{\mu\nu}$ or gravitons, which would thus appear as gravity
analogues of the Higgs boson in particle theory.

In the particle theory on the left-hand side, we recall that 
the self-interaction potential $V[\Phi]$ of the framons $\Phi$
plays an important role.  What happens if we introduce on the
right for gravity also a self-interaction potential for the
vierbeins?  The same symmetry arguments as before will lead to
a potential formally identical to that for the $su(3)$ theory of
the preceding section.  Now we noted in the analysis there that
the $su(3)$ framons remain orthonormal at vacuum corresponding
to a Euclidean metric in $\widetilde{su}(3)$ space until the
symmetry is broken by interaction with framons from $su(2)$.
Virtually the same analysis on the potential for the vierbeins,
i.e.\ the $H-\tilde{H}$ framons on the right-hand side of Fig.\  
\ref{Plan2}, will show that the vierbeins here will also be 
orthonormal at a stationary point of the potential (although
with the indefinite signature, it is here not a minimum as it was
for the other case).  This would seem to mean that when left to 
themselves, the vierbeins would settle to orthonormality and the
metric to the Minkowski metric.  In other words, in the absence of
or far away from other fields like matter, $X$ would settle down
automatically to a Minkowski world, which is probably the sort
of solutions we would seek in any case.  In regions of space
dominated by matter, the effects of the framon potential will
presumably be negligible.

Matter fields in the particle theory on the left are usually 
fermionic; fermionic fields too can be introduced on the right
for gravity with the ECKS formalism \cite{ecks} by means of the 
vierbeins
and the spin connection.  In either case, as usually formulated,
the fermion fields are inserted by hand without being ascribed
a geometrical significance.  In the framework of framed gauge 
theory we are considering, however, it appears that a possible
geometrical meaning for fermionic fields may emerge in the
following manner.

The symmetry displayed in Fig.\ \ref{Plan2} between the internal
structures on the left and the external structures on the right
may still seem defective in that the framons on the left, as
introduced in the preceding sections, are complex, whereas the 
vierbeins on the right are real.  This apparent asymmetry is 
due, however, only to our own sloppy convention.  When we have
spinor fields around, the symmetry group is not $SO(3)$  but
its double cover $SU(2)$.  In the same way, when we have spinors
on the right-hand side of Fig.\ \ref{Plan2}, as indeed there are
in nature, the symmetry groups $H$ and $\tilde{H}$ should not 
be taken as the Lorentz group $SO(3,1)$ as we did, but its
double cover $SL(2,\bbc)$\footnote{Strictly speaking, $SL(2,\bbc)$
  double covers the proper orthochronous Lorentz group, usually
  denoted $SO^+(3,1)$}.  That being the case, we ought to have
chosen as framons, in parallel to the framons in $\Phi$ of the 
preceding sections, not the vierbeins $e^a_\mu$ as we did, but 
the elements of a matrix, $\Psi$ say, whose rows transform as
4-spinors under $H$ and whose columns transform as 4-spinors
under $\tilde{H}$, removing thus what had appeared as a defect
before in the symmetry between the left and right side of Fig.\  
\ref{Plan2}. 
 
One interesting consequence of taking $\Psi$ as framons instead
of the vierbeins as one did above is to have space-time spinors 
now appearing as geometric objects.  However, since they transform 
as spinors under $H$, and $H$ is supposed to be confining, they 
will have to be confined and cannot be taken as actual fermion 
fields.  They can form bound states with their own conjugates 
giving bosonic fields like the graviton above, but can they form 
bound states with something else to give fermionic fields?  One 
possibility is their bound states with differential forms in $H$ 
space.  Saturating the $H$ indices in the framons with those of 
the differential forms will give objects which now carry only 
tilde indices, i.e.\ are neutral under $H$ and hence can propagate, 
but still transform as spinors under $\tilde{H}$ as fermion fields 
should.  But further, being differential forms, they will mutually 
anti-commute, again as fermion fields should.  Has one then found 
a geometric significance for fermionic matter?  
In other words, one says that they 
transform as spinors because they are framons in a theory with 
the double cover of the Lorentz group as symmetry group, and they 
anti-commute because they are basically line-elements.  Amusingly, 
a proton made up of 3 quarks will now appear as a volume element!  
We notice, though, that to saturate the spinorial indices in 
$\Psi$ one needs differential forms carrying also spinorial 
indices, i.e.\ differential forms not in ordinary space-time $X$ 
with space-time vectors $x,t$ as co-ordinates, but in a spinor-valued
version of it with details yet to be worked out.

By this stage, it is probably clear to all our readers that we are 
getting rapidly out of our depths, and that we have speculated
already much more than enough.  But these speculations have given
us at least some delectable food for thought and, together with
what the framed gauge theory framework seemed to have done above 
for the particle physics sector, new encouragement to continue 
with its exploration.

\end{document}